\documentclass{aa}  
\usepackage[varg]{txfonts}
\usepackage{graphicx}

\usepackage{amsmath}
\usepackage{txfonts}
\usepackage{natbib}
\bibpunct{(}{)}{;}{a}{}{,} 

\usepackage{xcolor}

\begin{document} 

\title{The influence of individual stars on the long-term dynamics of comets C/2014~UN$_{271}$ and C/2017~K2.
}

\titlerunning{The influence of stars on the dynamics of C/2014~UN$_{271}$ and C/2017~K2.}

\authorrunning{P.A.Dybczyński \& M.Królikowska}

\author{Piotr A. Dybczyński
          \inst{1}\fnmsep\thanks{\email{dybol@amu.edu.pl}}
          \and
        Małgorzata Królikowska \inst{2}\fnmsep\thanks{\email{mkr@cbk.waw.pl}}
        }

   \institute{Astronomical Observatory Institute, Faculty of Physics, Adam Mickiewicz University, Słoneczna 36, PL60-283 Poznań, Poland
   \and
   Centrum Badań Kosmicznych Polskiej Akademii Nauk (CBK PAN),
Bartycka 18A, 00-716 Warszawa, Poland}

   \date{Received XXXXXX; accepted XXXXX}

  \abstract
   {In June 2021 the discovery of an unusual comet C/2014~UN$_{271}$   (Bernardinelli-Bernstein) was announced. Its cometary activity beyond Uranus orbit   has also refreshed interest in similar objects, including C/2017K2   (PANSTARRS). Another peculiarity of these objects is the long interval of positional data, taken at large heliocentric distances. }
   { These two comets are suitable candidates for a detailed investigation of their long-term motion outside the planetary zone. Using the carefully selected orbital solutions, we aim   to estimate the orbital parameters of their orbits at the previous perihelion passage. This might allow us discriminating between dynamically old and new comets.}
   {To follow the dynamical evolution of long-period comets far outside the planetary zone, it is necessary to take into account both the perturbation caused by the overall Galactic gravitational potential and the actions of individual stars appearing in the solar neighborhood. To this aim, we applied the recently published methods based on stellar perturbers ephemerides.}
   {For C/2014~UN$_{271}$ we obtained a precise orbital solution that can be propagated to the past and to the future. For C/2017~K2 we have to limit ourselves to study only the past motion since some signs of nongravitational effects can be found in recent positional observations. Therefore, we use a specially selected orbital solution suitable for past motion studies. Using these starting orbits, we propagated both comets to their previous perihelia. We also investigated the future motion of C/2014~UN$_{271}$.}
   {Orbital evolution of these two comets appears to be sensitive to perturbations from several stars that closely approach the Sun. Unfortunately, some of these stars have 6D data with the uncertainties too large to obtain definitive results for the studied comets; however, it appears that both comets were probably outside the planetary zone in the previous perihelion.} 


   \maketitle

\section{Introduction}

We present the long-term dynamical evolution of two unusual long-period comets (hereafter LPCs): C/2014~UN$_{271}$ (Bernardinelli-Bernstein) and C/2017~K2 (PANSTARRS). We follow the abbreviation convention proposed by \cite{B-B:2021} and  for the brevity call these comets BB and K2, respectively.  In both cases, the first observations were taken at extremely large heliocentric distances. In addition, both comets attracted scientific attention because of their pronounced cometary activity well beyond Saturn's orbit. 

The discovery of BB was announced in June 2021 by \cite{BB:2021CBET} on the basis of  astrometric observations spanning 20 nights from 2014 Oct. 20 -- 2018 Nov. 8  (29.0~au -- 23.7~au from the Sun). A little later,  at a distance of  over 20~au from the Sun, observers reported the cometary activity of BB  \citep{Farnham:2021, Kokotanekova:2021, B-B:2021}. In all these papers, the authors estimate that BB might be one of the largest comets ever discovered. \cite{ B-B:2021} backward integration of the BB orbit under their assumed Galactic tidal model, and including perturbations from 8 stars taken from a list of Sun--star encounters published by \cite{Bailer-Jones:2018}, yields a perihelion distance of $\sim$18~au  during its previous perihelion passage 3.5~Myr ago. 

The  comet K2 observational campaigns have been conducted from the moment of its discovery in May 2017. Undoubtedly, its activity, most of all related to the existence of CO ice, was found at heliocentric distances well beyond 20~au \citep{Meech:2017, jewitt:2021_distant_activity_2017k2, Yang_Jewitt:2021}.

BB will pass through perihelion (10.95~au from the Sun) on January 22, 2031, whereas  K2 will reach its perihelion on December 19, 2022, at 1.80~au from the Sun. Therefore, both have been observed for many years on their pre-perihelion orbital leg, and in addition, at great distances from the Sun.  As a result, they are favorable candidates for studying their past motion and their origin. Moreover, such a large perihelion distance of BB allows us to make some statements on the future orbital evolution of this comet on the basis of pre-perihelion data because we can assume that nongravitational (hereafter NG) forces will be negligible in the context of the orbital motion of this comet.

Both comets have an original barycentric semimajor axis greater than 10\,000~au, so they belong to the so-called Oort spike. It is widely accepted that  it is necessary to take into account both Galactic and stellar perturbations when investigating the long-term dynamical evolution of so much elongated orbits, far outside our planetary system. A new, reliable, precise, and fast method of calculating the effect of these perturbations on LPC motion has recently been proposed by \citet{Dyb-Breiter:2021}. We use these methods in what follows. 

After applying the methods quoted above, we realized that increasing knowledge on stars visiting the solar neighborhood becomes more important in LPCs long-term dynamical studies. In recent years we have benefited  from the {\it Gaia} astrometric mission \citep{Gaia_mission:2016}. Using its latest third data release \citep{GaiaEDR3-summary:2021}, we show the effect of some particularly important stars passing near the Sun on BB and K2 motion. We also present the impact of the stellar data uncertainty on our results. 

\begin{table*}
\caption{\label{tab:positional-data} Characteristics of the positional data for both comets analyzed in this paper and the original $1/a$ obtained using an osculating orbit determined from this data arc. }
\setlength{\tabcolsep}{4.0pt} 
\centering
		\begin{tabular}{llcccccrcccr}
\hline \hline
Comet     & $q$    & first obs.   & last obs. &  $T$  & data & range of    & No   & Model  & $1/a_{\rm ori}$ & Ref & Model \\
          &        & \multicolumn{3}{c}{}             & arc  & helio-dist. & of   & of    & &      & name\\
          & [au]   & \multicolumn{3}{c}{[ yyyy\,mm\,dd ]} & [yr] &  [au]   & obs. & motion &[$10^{-6}$au$^{-1}$]  & &     \\
\hline
\\
C/2014 UN$_{271}$ & 10.9 & 2010\,11\,15 & 2021\,06\,22 & 2031\,01\,22 & 11.6 & 34.1~~--~~20.2   &   59 & GR & 50.17 $\pm$ 0.75 & here &  i1   \\
                  &      &              & 2021\,09\,19 &              & 11.8 & 34.1~~--~~19.9  &  114 & GR &  51.15 $\pm$ 0.56 & here & b8    \\
\\
C/2017 K2         & 1.81 & 2013\,05\,12 & 2018\,01\,23 & 2022\,12\,20 & 4.64 & 23.7~~--~~14.8  & 450  & GR & 48.48 $\pm$ 7.93 & KD18a & a8 \\
                  &      &              & 2018\,10\,13 &              & 5.42 & 23.7~~--~~13.0  & 996  & GR  & 35.49 $\pm$ 2.33 & KD18b & a9 \\
                  &      &              & 2020\,11\,13 &              & 7.51 & 23.7~~--~~7.81   & 3757 & GR & 33.02 $\pm$ 0.45 & here &  a6  \\
                  &      &              & 2020\,11\,13 &              & 7.51 & 23.7~~--~~7.81   & 3757 & NG & 35.97 $\pm$ 0.91 & here &  c5  \\
                  &      &              & 2021\,09\,18 &              & 8.35 & 23.7~~--~~5.31   & 6634 & GR & 42.78 $\pm$ 0.23 & here &  b5  \\
\hline                  
		\end{tabular}
		\tablefoot{
		The first two rows regarding K2 relate to the solutions discussed in \citet[][KD18a]{Kroli-Dyb:2018} and \citet[][KD18b, see Addendum]{Kroli-Dyb:2018c}, respectively.
		}
\end{table*}

In the abstract of their paper on K2, \citet{jewitt:2021_distant_activity_2017k2} wrote: {\it Nongravitational acceleration in C/2017~K2 and similarly distant comets, while presently unmeasured, may limit the accuracy with which we can infer the properties of the Oort cloud from the orbits of long-period comets}.
 However, as we discussed in \cite{kroli-dyb:2012}, uncertainties related to the NG~acceleration in the comet's motion can be eliminated or substantially minimized by using the pre-perihelion part of the data to calculate the osculating orbit. An even better effect will be when the pre-perihelion part of data is limited to large heliocentric distances. Then the original orbits will also not be burdened by the uncertainty related to the subsequent orbital change under the influence of increasing NG forces as the comet approaches the Sun.

Comets BB and K2 are  ideal for such research based on distant data before perihelion. Astrometric measurements starting from distances exceeding 20~au are available for both. It guarantees that their distant parts of data arcs are long enough to determine orbits of the highest quality class  \citep{Kroli-Dyb:2018b}. One of the goals of this study is to show how good quality orbits we can deal with by limiting the data arcs for orbit determination to the properly selected part of pre-perihelion orbit. Another goal is to verify how a particular  comet's past motion fits the general picture of cometary reservoir evolution; see \citet{Vokrouhlicky:2019} and  \citet{Dones:2015} for the extensive reviews. 

The structure of this paper is as follows.  In Sect.~\ref{sec:starting-orbits}, we discuss the choice of appropriate data arcs to obtain the starting orbits to study the origin of both comets. Our current state of knowledge on the potential stellar perturbers of LPCs motion is briefly described in Sect.~\ref{sect:stars} while  Sect.~\ref{sect:uncertainties} shows the methods of dealing with the orbital parameter uncertainties of comets and stars in our long-term dynamical studies. We also present data on all particular stars mentioned in this paper. Section~\ref{BB-past-evol} describes BB past orbit evolution and its parameters at the previous perihelion, while Sect.~\ref{sect:BB-future-evol} offers the same for the future BB dynamics. K2 past orbit changes , we discuss in detail in Sect.~\ref{sect:K2-past-evol}. Section~\ref{sect:summary} contains  a summary and our conclusions.

\section{Orbit determination for both comets\label{sec:starting-orbits}}

  We present a detailed description of the positional data used in this paper for both comets in Table~\ref{tab:positional-data}. This table also includes values of an original $1/a$.  The model of motion and the methods used for osculating orbit determination are described in \cite{kroli_dyb:2017} and references therein. 
 
\subsection{The orbit of BB}
At the end of September 2021, comet BB was about 20~au from the Sun, and we took the entire set of astrometric data available at the IAU Minor Planet  Center\footnote{https://www.minorplanetcenter.net/db\_search/}.  We supplemented this data with the single precovery positional detection found by \citet{B-B:2021}. Due to this single precovery measurement, the data arc of BB  has extended by about 3.7~yr, going back to 2010 November~15. Additionally, \citet{B-B:2021}  suggested some corrections of positional data, and we applied them according to Table~2  in their paper. It turned out that the precovery measurement and the corrections to other positions resulted in only a slight change in the orbital elements; for example, the value of original $1/a$ changed at  a level below 1\%. We conclude that the orbit of BB is now quite well constrained.	
 
As expected, we found no evidence of NG effects in the motion of BB, which is still beyond the orbit of Saturn. 
Table~\ref{tab:positional-data} also shows the solution 'i1' based on a 3-month shorter data arc and almost two times smaller number of positional measurements than the data arc used in the case of the 'b8' solution. Both $1/a_{\rm ori}$ values  do not differ in statistical terms (at one sigma level). Furthermore, based on the data arc almost a year shorter than that used for solution 'i1' \citet{B-B:2021} obtained the value of 49.50$\times 10^{-6}$~au$^{-1}$ for $1/a_{\rm ori}$, which indicates the high compatibility of the $1/a_{\rm ori}$ as a function of the increasing data arc taken for the orbit determination.

\subsection{The orbit of K2 for the backward dynamical evolution}

 For K2, the situation is more interesting regarding the NG~effects and is, therefore, more complex from the perspective of the past dynamical evolution study.
Our first published attempt to study the past dynamics of K2 was based on  the 4.6-yr data arc  \citep[][hereafter KD18a]{Kroli-Dyb:2018}.  Almost a year later, we updated our calculations based on a longer interval of observations and published our results as an addendum to our preprint  available at arXiv \citet[][hereafter KD18b]{Kroli-Dyb:2018c}. These two orbital solutions were presented as Solution A1 in KD18a and Solution A1-new in the  addendum of KD18b . In the present paper we call them 'a8' and 'a9', respectively, see Table~\ref{tab:positional-data}.  Here, we analyze whether a longer observational arcs will improve the reliability of the original K2 orbit. We decided to include in Table~\ref{tab:positional-data} three additional solutions, 'a6', 'c5', and 'b5', based on longer data arcs. 

It turned out that currently some traces of NG~effects can be seen in the longer data arcs considered here; see solution 'c5'. For this  orbit, we obtained  the following NG~parameters: $A_1 = 28.414 \pm 4.968$, $A_2 = 11.610 \pm 1.089$, and $A_3 = 4.5795 \pm 0.7221$, in units of $10^{-8}$day~au$^{-2}$, where g(r)-like function reflects CO-driven sublimation and was applied as described in \cite{krolikowska:2020}. Fortunately, we found that this NG~orbit gives a very similar original $1/a_{\rm ori}$ as in the purely gravitational case 'a6', based on the same interval.  In addition, the NG~solution for the longest data arc (the same as in the case of the GR~solution marked as 'b5') results in  a  small and uncertain radial component of NG~acceleration (using the CO-driven formula) and allowing for NG effects does not eliminate some trends visible in [O-C], so it is not considered here. By  the analysis of many NG and GR orbits, we finally concluded that the best solution for studying long-term past dynamics of K2 is the solution a9. Thus, solutions based on longer data arcs ('a6', 'c5', and 'b5') are only used here as a check to  which extent longer data arcs will change our inference about the dynamical status of K2.  Choosing 'a9' solution  is a rather cautious approach, because the uncertainty of $1/a_{\rm ori}$ is two times greater than for the case of 'c5' solution.

\subsection{Future orbital evolution predictions}

Today we do not know how the NG~acceleration will change the current orbit of K2 when a comet pass the perihelion at a distance of 1.8~au from the Sun in December 2022. Therefore, its future dynamics can only be discussed when the comet  will be on its post-perihelion trajectory. The situation is different in the case of BB, because its perihelion distance equals almost 11~au. Thus, it seems that today's predictions will be correct also in the aspect of a future dynamical evolution for this comet. 

\section{Stellar perturbers}
\label{sect:stars}
Our knowledge on stars that can perturb LPCs motion has expanded in the last years. In \cite{rita-pad-magda:2020} they have introduced a publicly available StePPeD database\footnote{\url{https://pad2.astro.amu.edu.pl/StePPeD}} containing data on potential stellar perturbers of  LPCs motion. This database has been revised several times in 2020, which  finally led to StePPeD release 2.3, mostly based on the {\it Gaia} DR2 catalog \citep{Gaia-DR2:2018} . However, after  the {\it Gaia} Early Data Release 3 has been made available \citep{GaiaEDR3-summary:2021}  it became clear that the next substantial update of the StePPeD database was necessary.

The first step of this upgrade has been finished in September 2021, and a new StePPeD release 3.0 was announced. This first step  was limited to the same list of stars as in version 2.3, but for almost all  of them, the new data from  the {\it Gaia} EDR3 catalog  have  been incorporated.  This work is in progress to add new stellar perturber candidates. 

We note here that a comparison of  the data between DR2 and EDR3  sometimes revealed substantial changes. Almost 25\% of the stars in our list have a new parallax that differs by more that 100\% when compared to its previous (i.e., DR2) value; nine of them have a negative parallax in EDR3. Over one-third of the stars in our database have the RUWE  parameter greater than 1.4, which according to the {\it Gaia} documentation indicates possibly less accurate data \citep{Gaia-EDR3-validation}. For five percent of our stars there are no parallax neither proper motion in EDR3 and we still have to rely on the {\it Gaia} DR2 data.

In November 2021, radial velocities of a dozen of stars have been updated, resulting in the release 3.1 of StePPeD database (see footnote 1), for details see the 'Changelog' available at the StePPeD Internet page. All calculations presented in this study are based on this latest version of the StePPeD database.

\begin{table*}
	\caption{ List  of stars mentioned in this paper with their parameters of the closest approach to the Sun. \label{tab:star-list} }
    \centering
    \begingroup
    \renewcommand{\arraystretch}{1.7} 

\begin{tabular}{c c l c r c c c} 
		\hline \hline
        \multicolumn{3}{c}{S t a r ~~i d e n t i f i c a t i o n} & \multicolumn{4}{c}{P a r a m e t e r s ~~o f ~~t h e ~~a p p r o a c h ~~t o ~~t h e ~~S u n} &   \\	
		StePPeD & Common & \multicolumn{1}{c}{{\it Gaia} EDR3} &  & \multicolumn{1}{c}{Distance statistics} & & relvel & Mass \\
		ID & name & \multicolumn{1}{c}{ID} & mindist [pc] & p05 - median - p95 & mintime [Myr] & [kms$^{-1}]$ & [M$_{\odot}$]
		\\
		\hline
		P0107 & \object{Gliese 710} & 4270814637616488064 & $0.052\pm0.005$ & 0.048--0.052--0.056 & $+1.290\pm0.001$ & 14.8 & 0.65 \\
		P0111 & \object{HIP 94512} & 4306481867124380672 & $1.037\pm0.084$ & 0.969--1.037--1.108 & $+3.30\pm0.04$ & 31.1 & 1.69 \\
		P0230 & \object{HD 7977} & 510911618569239040 & $0.014 \pm 0.057$ & 0.008--0.032--0.071& $-2.47 \pm 0.03$ & 30.7 & 1.08 \\
        P0417 & \object{Ton 214}        & 1281410781322153216  & $0.514\pm0.025$ & 0.492--0.514--0.535 & $-1.47\pm0.01$ & 32.7 & 0.85 \\
		P0506 &                & 5571232118090082816  & $0.199\pm0.012$ & 0.189--0.199--0.208& $-1.08\pm0.01$ & 90.2 & 0.77 \\
		P0508 & & 2946037094762449664 & $0.263\pm0.546$ & 0.113--0.382--0.755 & $-0.98\:^{+0.21}_{-0.18}$ & 39.9 & 0.25 \\
		P0509 &                &   52952720512121856  &  $0.318\pm0.223$ & 0.176--0.333--0.501 & $-0.67\:^{+0.05}_{-0.06}$ & 31.3 & 1.46 \\
        \hline
	\end{tabular}
	\endgroup
	\tablefoot{
The first three columns show the star identifiers, the four next  columns present parameters of the closest approach to the Sun, and the last column contains a star mass estimate.
	}
\end{table*}

\begin{figure}
	\includegraphics[angle=270,width=1\columnwidth]{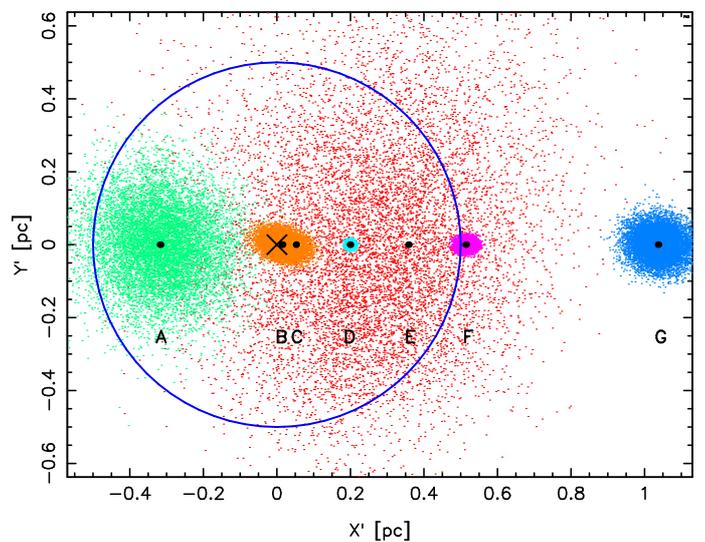} 
	\caption{ Seven stars mentioned in this paper --  the comparison of their nominal minimal distances from the Sun and the spread of star clones at their closest approach. Black dots in the center of each clone cloud mark the nominal star position at the closest Sun--star approach;  see text on how this composite figure was obtained. All 10\,000 clones of star C are hidden under its black dot since the stellar data are extremely precise in this case. Label meanings: A: P0509, B: P0230 (\object{HD~7977}), C: P0107 (\object{Gliese~710}), D: P0506, E: P0508, F: P0417 (\object{Ton~214}), G: P0111 (\object{HIP~94512}) ; see Table~\ref{tab:star-list} for details on each star.  The blue circle marks the usually adopted outer limit of the Oort cloud.}\label{fig-seven-stars}
\end{figure}

During the study of  the long-term dynamical evolution of BB and K2, we recognized very strong perturbations caused by stars \object{HD~7977} and \object{Gliese~710} (named P0230 and P0107 in the StePPeD database, respectively) and weaker but potentially important perturbations from five other stars. The details for all seven stars are shown in Table~\ref{tab:star-list}, including their identifiers, parameters of the closest approach to the Sun, and mass estimates.  We describe the minimal Sun--star distance in two different ways.  The value presented in the fourth column, 'mindist', is calculated in a special way: this is the distance from the Sun to the centroid of a cloud of 10\,000 star clones drawn according to the data uncertainties, using the respective covariance matrix.  As the uncertainty  of the mindist parameter, we present  here a radius of a sphere around that centroid, which includes 90 percent of star clones.  Uncertainties greater than the mindist directly indicate that the clone cloud surrounds the Sun (see Col.~4 for P0230 star).  In the fifth column, we present a formal statistical description of the Sun-star distance set for all stellar clones using three percentiles: 5\% (p05), the median and 95\% (p95). They are also expressed in parsecs.

The comparison of the stellar data uncertainties and their minimal distances to the Sun, we present graphically in Fig.~\ref{fig-seven-stars}. For each star listed in Table~\ref{tab:star-list}, we plot positions of its 10\,000 clones at their closest approach to the Sun, projected on the plane of the maximum scatter; for details of such calculations, see \cite{dyb-berski:2015}. The  seven different plots of the spread of star clones were then merged into a single image.  Centroids of all clouds, are aligned along the horizontal line keeping the correct distance from the Sun and maintaining the same scale in clone clouds spread. The whole clone cloud of P0107 (C in Fig.~\ref{fig-seven-stars}) is hidden under its centroid black dot due to extremely precise data for this star.  It should be stressed that the seven stellar close approaches to the Sun presented in Fig.~\ref{fig-seven-stars} happened or will happen in different epochs spread over 6~Myr interval; see the  sixth column of Table~\ref{tab:star-list}.

There is an important qualitative difference between the perturbative action of P0230 and P0107 and the remaining five stars mentioned above.  According to the most recent data, approaches of these two stars to the Sun are so close that they noticeably perturb the solar galactic trajectory. As a result, these perturbations impart all Solar System bodies' motion. The resulting heliocentric orbit changes depend at most on a body's heliocentric velocity. LPCs, when far from their perihelion and therefore moving very slowly, are the best candidates for the largest change in their perihelion distance. We show this effect in the following sections.

\begin{table*}
		\caption{\label{tab:BB_previous_next} Orbital parameters of BB at the previous and next perihelia for different types of simulations based on BB orbit solutions 'i1' and 'b8' given in Table~\ref{tab:positional-data}; each value is a median or mean  of  a set of thousands of clones used in a particular simulation with an uncertainty resulting from the obtained distribution. In all cases, perturbations by full Galactic tides are included, while the stellar perturbations are included as follows. For the past motion -- BB-prev-A: swarm of 5001 clones of BB, without stellar perturbation;  BB-prev-B:  swarm of 5001 clones of BB, perturbations by all stars included (using nominal orbits); BB-prev-C: nominal orbit of BB and 10\,000 clones of star P0230; BB-next-D: 10\,000 pairs: comet clone and P0230 clone. For the future motion -- BB-next-A: swarm of 5001 clones of BB, without stellar perturbation; BB-next-B: swarm of 5001 clones of BB, perturbations by all stars included (using nominal orbits); BB-next-C: nominal orbit of BB and 10\,000 clones of star P0107; BB-next-D: 10\,000 pairs: comet clone and P0107 clone. 
		} 
		\setlength{\tabcolsep}{8.0pt} 
\centering
		\begin{tabular}{ccccccc}
			\hline \hline 
simulation  & \multicolumn{3}{c}{ s o l u t i o n ~~~i1} & \multicolumn{3}{c}{s o l u t i o n ~~~b8}      \\
 series     & $q_{\rm prev}$ & $1/a_{\rm prev}$ & $P_{\rm prev}$ & $q_{\rm prev}$ & $1/a_{\rm prev}$ & $P_{\rm prev}$ \\
            & [au] & [$10^{-6}{\rm au}^{-1}$] & [Myr] & [au] & [$10^{-6}{\rm au}^{-1}$] & [Myr] \\  \hline
\\
& \multicolumn{6}{c}{BB ~~a t ~~~p r e v i o u s ~~~p e r i h e l i o n }\\
BB-prev-A & 14.99--15.29--15.62 & $50.15 \pm 0.75$   & 2.81 & 14.77--14.98--15.20 & $51.15 \pm 0.56$    &  2.73 \\
BB-prev-B & $294.7 \pm 43.0$    & 44.65--44.88-44.94 & 3.32 & $237.6 \pm 32.9$ & 44.36--44.81--44.93 &  3.33 \\
BB-prev-C & --                  &  --                &--& 12.28--50.15--254.1 & 43.97--49.92--52.34 &  2.83 \\
BB-prev-D & --                  &  --                &--& 12.50--49.62--252.4 & 44.10--49.84--52.46  & 2.84 \\
			\\
& \multicolumn{6}{c}{BB ~~~a t ~~~n e x t  ~~~p e r i h e l i o n}\\
BB-next-A & 2.264--2.619--3.004 & $36.07 \pm 0.75$ & 4.61 & 2.727--3.015--3.317 & $37.06 \pm 0.56$ & 4.43 \\
BB-next-B & $10.33 \pm 0.60$    & $35.12 \pm 0.78$ & 4.80 & $11.11 \pm 0.42$    & $36.16 \pm 0.59$ & 4.60 \\
BB-next-C &--                   &  --              &--& 9.23--11.08--13.45  & 36.09--36.17--36.23 & 4.59 \\
BB-next-D &--                   &  --              &--& 9.13--11.06--13.50  & 35.40--36.16--36.90 &  4.60 \\
			\hline
		\end{tabular}
\end{table*}

\begin{table*}
		\caption{\label{tab:K2_previous_part1} Influence of the K2 orbit uncertainties on the elements at the previous perihelion. For each orbital solution, we use 5001 clones of K2 orbit. Simulation K2-prev-A describes the results obtained when all stellar perturbations were excluded, while K2-prev-B reflects the results obtained from the full dynamical model with 232 potential stellar perturbers included, using their nominal tracks. In both variants the full Galactic potential was taken into account. 
		}		
		\setlength{\tabcolsep}{6.0pt} 
\centering
		\begin{tabular}{lccccc}
			\hline \hline
solution         & a8                 & a9                   & a6                 & c5                  & b5 \\   \hline
\\
\multicolumn{6}{c}{K2-prev-A:} \\
$q_{\rm prev}$ ~~~~[au]  & 2.751--3.999-8.755 & 8.449--12.00--18.13 & 15.47--16.87--18.48 &	9.815--11.24--13.06 & $5.691 \pm 0.101$ \\
$1/a_{\rm prev}$ [$10^{-6}{\rm au}^{-1}$] & $48.26 \pm 7.93$   & $35.48 \pm 2.33$     &$33.01 \pm 0.46$    &$35.96 \pm 0.91$   &$42.77 \pm 0.23$ \\
$P_{\rm prev}$ ~~~~[Myr]    & 2.98  & 4.73 & 5.27 & 4.63  &  3.57 \\
&&&&&			\\
\multicolumn{6}{c}{K2-prev-B:}\\
$q_{\rm prev}$ ~~~~[au]	 & 1.790--108.5-380.2 & 373.5--441.6--490.0 & 474.9--483.7--491.5 &	406.1--431.1--454.1 & $257.4 \pm 6.3$ \\
$1/a_{\rm prev}$ [$10^{-6}{\rm au}^{-1}$] & 37.84--49.79--58.31& $34.88 \pm 2.61$     &$32.12 \pm 0.50$    &$35.41 \pm 1.02$   &$43.19 \pm 0.27$ \\
 $P_{\rm prev}$ ~~~~[Myr]    & 2.84  &  4.85  & 5.49 &  4.74 &  3.52 \\
\hline
		\end{tabular} 
\end{table*}

\begin{table*}
		\caption{\label{tab:K2_previous_part2} Impact of uncertainties on the previous  K2 orbit elements obtained from three different simulations. For K2-prev-C we used the nominal K2 orbit and 10\,000 clones of P0509. In K2-prev-D, a similar model was used but with 10\,000 clones of P0230. K2-prev-E consisted of 10\,000 pairs:  one P0230 clone and  a random K2 clone from our set of 5001 ones. Full Galactic potential and all remaining stars on their nominal tracks were taken into account. All these numerical experiments were based  on the 'a9' orbital solution of K2. 
}	
		\setlength{\tabcolsep}{12.0pt} 
\centering
		\begin{tabular}{lccc}
			\hline \hline
simulation  &  &  &  \\
 series     & \multicolumn{1}{c}{K2-prev-C} & \multicolumn{1}{c}{K2-prev-D }   & \multicolumn{1}{c}{K2-prev-E}   \\   \hline
\\
 $q_{\rm prev}$	~~~~[au]    & 407.1--447.6--489.1 & 5.299--147.3--1064.  & 5.163--144.0--1065. \\
 $1/a_{\rm prev}$ [$10^{-6}{\rm au}^{-1}$]  & 34.69--35.03--35.27 &  34.42--35.07--35.16 & 31.65--34.82--38.02 \\
 $P_{\rm prev}$ ~~~~[Myr]    & 4.82  &  4.81  & 4.86 \\
			\hline
		\end{tabular}
\end{table*}

\section{Dealing with the uncertainties}\label{sect:uncertainties}

In the next three sections, we describe in detail the past and future motion of BB and the past motion of K2. For both comets, we also estimated the influence of their orbital uncertainty on a previous or next orbit for all solutions presented in Table~\ref{tab:positional-data}.  In addition, for the preferred solutions, we performed simulations to observe the effects of stellar uncertainties in various cases. To this purpose, we extensively use the methods proposed recently by \cite{Dyb-Breiter:2021} and the stellar data from the latest release 3.1 of the StePPeD database.

In determining an osculating orbit, we obtain the covariance matrix that allows us to construct comet orbit clones satisfying the observational constraints.

Dealing with the cometary orbit uncertainty is fairly  straightforward. During an osculating orbit determination we obtain the covariance matrix  that allows us to draw comet orbit clones satisfying the observational  constraints, details  of this procedure  can be found in \citep{sitarski:1998}. In all simulations presented in this paper, we use 5000~comet clones plus a nominal orbit.  Each clone in this set, we numerically propagated back and forth to obtain original and future swarms of barycentric orbit clones. Investigating the previous (or next) orbit, we repeat the backward (or forward) numerical integration for each comet clone exactly as described in \cite{Dyb-Breiter:2021}, using the  latest stellar ephemerides  from the StePPeD site. In all cases, we take into account both the overall Galactic gravitational potential and stellar perturbations from a set of 407 potential perturbers. Due to the usage of advanced algorithms,  this calculation is precise and very fast, for 5001 comet orbits, we obtain the results in 15--20~minutes on a standard PC. For the sake of comparison, we also present the results of calculations for comet clones with all stellar perturbations excluded.

Analyzing the effect of stellar data uncertainties requires a more complicated and much more time-consuming approach. Each individual calculation consists of two stages. First, we generate a tailored stellar ephemeris and then use a standard code (described above) for a comet motion propagation, using this individualized stellar ephemeris. To obtain this special ephemeris, we replace the nominal data for the studied star with its clone drawn from the respective 6-dimensional covariance matrix from {\it Gaia} EDR3\footnote{see the {\it Gaia} EDR3 documentation, paragraph 4.1.7.0.4 on page 196}. Then we generate the stellar ephemeris using nominal data for the remaining stars. In this way, we still take into account all possible star--star interactions, which  are very rare but happen. In all simulations presented in this paper, we use 10\,000 star clones for the star in question plus the nominal star tracks for the rest of them. The tailored stellar ephemeris can be produced for a shorter time than the publicly available one (30~Myr back and forth), but it still takes a considerable CPU time, typically about 20~secs. At the second stage, the subsequent comet motion integration takes only milliseconds, but a whole investigation consisting of 10\,000 cases takes over 50~hours on a single CPU core.

For calculating the effect of the stellar perturbation on a comet motion, we have to know the star mass.  In the StePPeD database, there are mass estimates included together with their uncertainties and we used these values. These data are taken from numerous different sources.  However, the uncertainties in the mass estimate were  obtained by various considerably different methods. Their nature is highly non-linear, often asymmetric, and with unknown error distribution. Therefore, the statistical interpretation of the mass estimate uncertainties is difficult, and we decided not to draw random mass values from these data, and in all calculations we utilize the nominal mass estimate. 

However, we should keep in mind that the mass uncertainty is an additional source of the approximate nature of our results. For example, the mass estimate for the most important star, P0230, is taken from \citet{TIC-8:2019} and is described as 1.080~M$_{\odot}$ with a symmetric uncertainty of $\pm$0.136~M$_{\odot}$. The nominal value of the  previous perihelion distance of K2 is 441~au (for the 'a9' solution) but if we use a lower limit mass for P0230, that is,  0.944~M$_{\odot}$, we will obtain  a lower value of $q_{\rm prev}=331$~au.

Using the above  procedure, we can theoretically investigate the simultaneous effect of two or more star uncertainties without additional computational cost, since the drawing a star clone is  very fast. However, the spread of the results considerably grows. Therefore, it is necessary to calculate a much larger number of cases to achieve a statistically valuable result. Fortunately, in all cases described in the following sections, only one star strongly perturbs a comet motion during the previous or next orbital period.

\begin{figure}
	\includegraphics[angle=270,width=1\columnwidth]{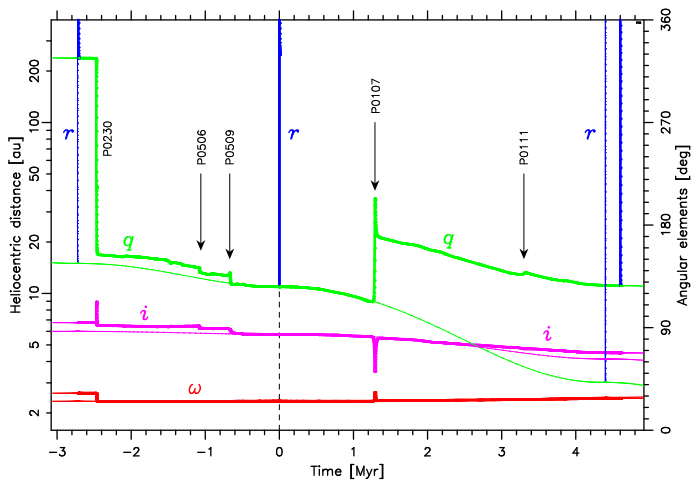}
	\caption{Past and future dynamical evolution of  C/2014~UN$_{271}$ nominal orbit (b8 solution). Changes in a perihelion distance (green), an inclination (fuchsia) and an argument of perihelion (red) are shown. The thick lines depict the result of the full dynamical model while thin lines show the evolution of elements in the absence of any stellar perturbations, i.e., only the Galactic perturbations are taken into account. Angular elements are expressed in a Galactic frame. We show also names of stars that make significant impart on this dynamical evolution.\label{fig:2014un_nomial}}
\end{figure}

\begin{figure}
	\includegraphics[angle=270,width=1\columnwidth]{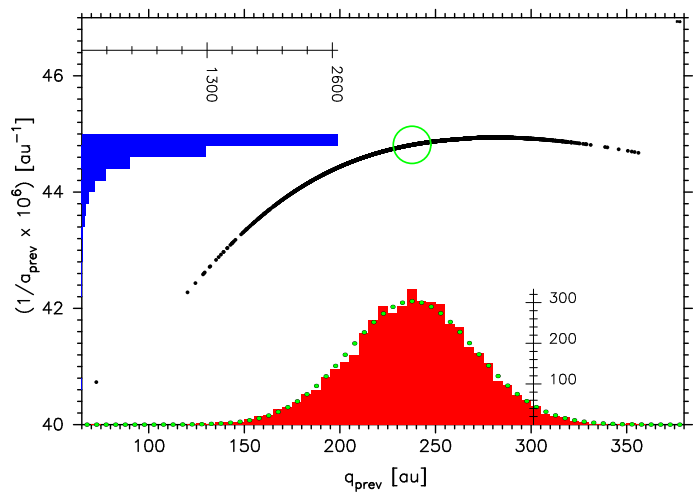} 
	\caption{ Influence of BB orbit uncertainty on parameters of the previous orbit. Black dots show the results of the backward numerical integration of 5001 clones of BB stopped at the previous perihelion with all Galactic and stellar perturbations included.  The center of the green circle marks the nominal result. The red histogram presents a marginal distribution of $q_{\rm prev}$ along with the best Gaussian approximation (green dots in the middle of each bar). The blue histogram is a marginal distribution of $1/a_{\rm prev}$.}\label{fig-2014unb8_hist}
\end{figure}

\begin{figure}
	\includegraphics[angle=270,width=1\columnwidth]{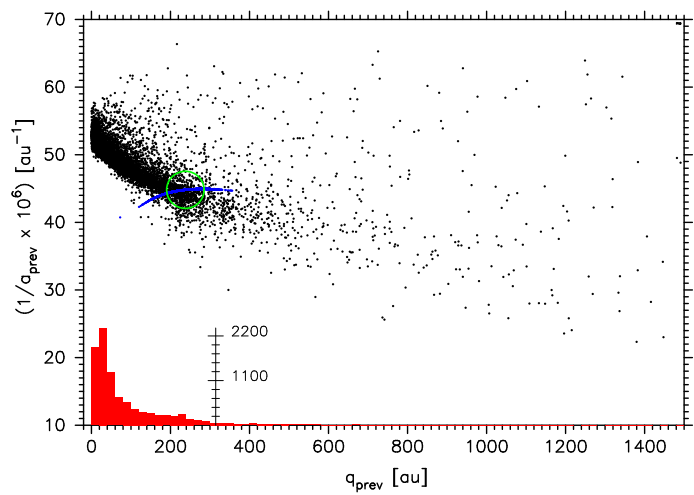} 
	\caption{ Influence of P0230 data uncertainty on parameters of the previous BB orbit -- in this simulation we used 10\,000 pairs: a comet clone and a P0230 clone. Each pair's result is represented by a black dot. Omitted are 114 points from a long right tail, among them 27 with negative $1/a_{\rm prev}$. The marginal distribution of $q_{\rm prev}$ is performed for all black points and shown as the red histogram below the main plot.  For the sake of comparison, the main result is overprinted with the blue swarm of dots showing the result of an additional experiment: 5001 BB orbits and nominal P0230 track. The nominal result is in the center of a green circle. }\label{fig-2014unb8_230summ}
\end{figure}

\section{C/2014 UN$_{271}$ past orbit evolution}\label{BB-past-evol}

Starting from the up-to-date osculating orbit of BB (solution b8 in Table \ref{tab:positional-data}) we calculated the original and future orbits at a distance well outside the planetary perturbation zone (as usual, we used a distance of 250~au from the Sun). The elements of these orbits are presented in Appendix~\ref{Ap:orbit_UN271} in Tables~\ref{tab:UN_original_orbits} and \ref{tab:UN_future_orbits}. Using these original and future orbital elements, we calculated the dynamical evolution of the nominal BB orbit for one orbital period to the past and to the future. This is presented in Fig.~\ref{fig:2014un_nomial}. In the past motion, we see several small perturbations from passing stars and one very strong orbit change. The two small interactions were with P0509 at $-0.67$~Myr and P0506 at $-1.08$~Myr, see Table \ref{tab:star-list} for more information on these stars. However, the most prominent perturbation was caused 2.47~Myr ago by the star P0230. As shown in Fig.~\ref{fig:2014un_nomial} the nominal previous perihelion distance equals 238~au and BB was at the previous perihelion nominally 2.71~Myr ago. If we exclude all stellar perturbations (the thin lines in Fig.~\ref{fig:2014un_nomial}), then the previous perihelion distance will be much smaller, only 15~au.

It should be stressed that the dynamical evolution depicted in Fig.~\ref{fig:2014un_nomial} is based on the nominal BB data and nominal data for all 232 stars included in the dynamical model of the past comet motion. To obtain a more realistic picture, we should estimate the influence of the cometary and stellar data uncertainties and add the results to the above discussion. In  the first step, we calculated the effect of the BB orbit uncertainty using the methods described in Sect.~\ref{sect:uncertainties}. The result of this numerical experiment is summarized in Fig.~\ref{fig-2014unb8_hist}. The previous perihelion distance appears to be very close to the nominal value and can be described as $237.6 \pm 32.9$~au since its distribution can be quite well approximated with the Gaussian one, see also  the BB-prev-B row in Tab.~\ref{tab:BB_previous_next}.

As is clearly shown in Fig.~\ref{fig:2014un_nomial}, apart from a series of weak stellar perturbations, the strongest one is caused by P0230. Its importance comes additionally from the fact that it is a perturbation of the Sun motion, so it acts on the comet indirectly and independently on the star--comet distance. Instead, it strongly depends on the minimal distance of the Sun--star encounter and its geometry. In Fig.~\ref{fig-seven-stars} we present the effect of P0230 uncertainties (Star B) on these parameters, drawing 10\,000 clones of this star and stopping their motion at the closest approach to the Sun. From this picture, it is obvious that both the minimum distance can be arbitrary small and the direction of the impulse imparted on the Sun is unknown.

To observe the possible effect of P0230 uncertainties on the past BB motion, we performed the calculation in which we used 10\,000 pairs:  a P0230 clone and a BB clone as described in Sect.~\ref{sect:uncertainties}.  The result is shown in Fig.~\ref{fig-2014unb8_230summ}. The spread of the previous orbit parameters is considerable, and to make this plot readable, we have to omit a very long tail of points placed to the right. This omitted set of points consists of 27 cases of negative $1/a_{\rm prev}$ and 87 large values for elliptic orbits. For the sake of comparison, we overprinted the main plot with the swarm of blue dots showing the distribution of the previous BB orbit when only the nominal star P0230 acts on 5001 comet clones. This blue set of points is the same as that shown in black in Fig.~\ref{fig-2014unb8_hist}.

The smallest $1/a_{\rm prev}$ value obtained in this numerical experiment equals $-558.7$ in the same units as in the plot. The interval of $q_{\rm prev}$ spreads from 0.021 to 20\,908.3~au. The distribution of $q_{\rm prev}$ is strongly non-Gaussian and we can describe it with three deciles, 10$^{\rm th}$, 50$^{\rm th}$ (median), and 90$^{\rm th}$: 12.5--49.6--252.4~au. 

In Fig.~\ref{fig-2014unb8_230summ} it can be seen  that the nominal $q_{\rm prev}=237.74$~au is far from the maximum of the $q_{\rm prev}$ distribution. The reason for this comes from the geometry of the cloud of P0230 clones with respect to the Sun (star B in Fig.~\ref{fig-seven-stars}). The nominal minimum distance between P0230 and the Sun is 0.014~pc. However, if we ignore the geometry and  analyze one-dimensional minimal distance distribution, the resulting median equals 0.032~au (see the fifth column in Table~\ref{tab:star-list}), and over 87\% of the values are greater than the nominal one. As a result, the much weaker perturbation producing  a smaller $q_{\rm prev}$ is much more probable.

The conclusion from this section is rather obvious: before much more precise data for P0230 are  provided, we will not be able to describe the BB orbit at its previous perihelion in a definitive manner. We look forward to the next {\it Gaia} data release,  which is announced to happen in June 2022.

\section{C/2014 UN$_{271}$ future orbit evolution}\label{sect:BB-future-evol}

\begin{figure}
	\includegraphics[angle=270,width=1\columnwidth]{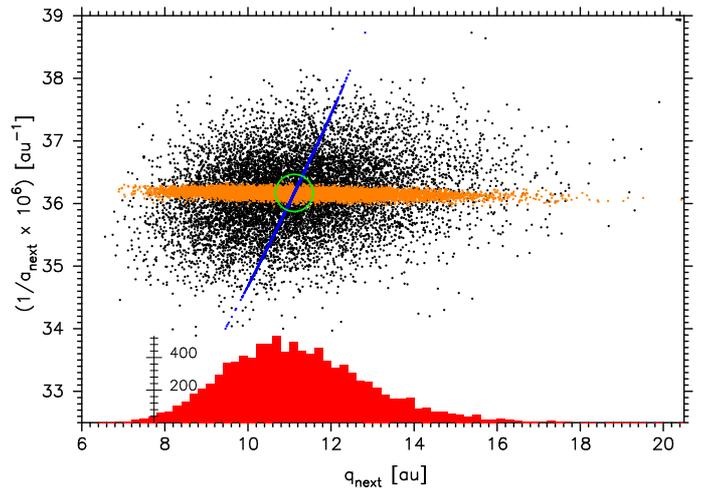} 
	\caption{ Composite picture of the influence of cometary and stellar uncertainties on the BB orbit recorded at the next perihelion passage. Blue dots mark 5001 clones of BB, each  perturbed by all stars on their nominal tracks. Orange dots represent the nominal BB orbit perturbed by 10\,000 clones of P0107. Black dots presents 10\,000 pairs of the interacting BB and P0107 clones. Red marginal distribution histogram of $q_{\rm next}$ corresponds to the black cloud of points. The result for nominal stellar and cometary data is in the center of the green circle.}\label{fig-2014unb8_107_3mix}
\end{figure}

In the future BB motion part of Fig.~\ref{fig:2014un_nomial} we can observe a strong perturbation caused by P0107 and a barely visible,  weaker one by P0111. Similarly to the past BB motion, the strongest perturbation is an indirect one, since it results from a very close passage of P0107 near the Sun in 1.29~Myr. However, in this favorable case, the prediction of the future is much easier than the prediction of the past . This comes from the very precise data available for P0107. As was already mentioned, in Fig.~\ref{fig-seven-stars} all the P0107 cloud of 10\,000 clones are hidden under the black dot which marks the  star's  nominal position at the closest encounter with the Sun.

Using the methods described in Sect.~\ref{sect:uncertainties} we performed three numerical simulations to observe the  effect of cometary and stellar data uncertainties on the BB next orbit parameters: BB-next-B -- all stars on their nominal tracks acting on each of the 5001 BB clones, BB-next-C -- the same star set but in each run P0107 is replaced by one of 10\,000 clones, acting on the nominal BB orbit, and BB-next-D -- 10\,000 pairs: a P0107 clone and a BB clone. The numerical results are presented in Table~\ref{tab:BB_previous_next}. Since the influence of cometary orbit uncertainties appeared comparable to that of stellar data, we summarized all three simulations in one composite plot shown in Fig.~\ref{fig-2014unb8_107_3mix}. We plotted the results of simulation BB-next-D as black dots overprinted with orange dots from BB-next-C, and the results of simulation BB-next-B as blue dots on top of the previous two sets. As it is clearly shown in Fig.~\ref{fig-2014unb8_107_3mix} stellar perturbations only slightly enlarge the interval of $q_{\rm next}$ values resulting from  the BB orbital uncertainty. As was stated at the end of sect.~\ref{sect:stars}, the perturbation of P0107 on BB is only indirect, resulting from an impulse gained by the Sun. It is well illustrated by the fact that in our BB-next-D simulation the smallest distance between BB clone and P0107 clone was over 55\,000~au.

For the future BB motion, our conclusion is that its next perihelion  could  be described by deciles (9.13--11.06--13.50)~au that is with a median value almost identical  to  the nominal one (11.11~au). The semi-major axis of the next BB orbit is larger than for the current apparition, mainly due to the planetary perturbations. Our simulation BB-next-D gives $1/a_{\rm next}$ = (35.40--36.16--36.90) $\times 10^{-6}$au$^{-1}$ which corresponds to the semimajor axis of about 27\,700~au and an orbital period of 4.6~Myr.

If one compares the original and future BB orbits (see Tables~\ref{tab:UN_original_orbits} and \ref{tab:UN_future_orbits}) it can be noticed that while planetary perturbations will change the semi-major axis of this comet orbit by almost 30\%, its perihelion distance will remain almost unchanged. After passing the planetary zone, the orbital period of BB will increase to  4.6~Myr, and, consequently, this comet will reappear among planets.

\section{Past dynamical evolution of C/2017 K2 orbit}\label{sect:K2-past-evol}

\begin{figure}
	\includegraphics[angle=270,width=1\columnwidth]{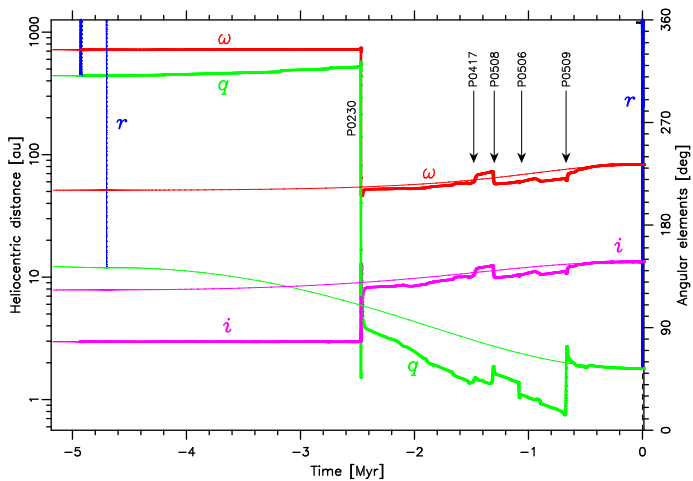}
	\caption{Dynamical past evolution of the nominal orbit of C/2017~K2 based on the 'a9' orbital solution. Several individual stellar perturbations are marked. The colors and line thickness meanings are the same as in Fig. \ref{fig:2014un_nomial} \label{fig:2017k2a9_nomial}}
\end{figure}

As was explained in detail in Sect.~\ref{sec:starting-orbits} we study only the past long-term dynamical evolution of K2 and base our calculations on the original orbit obtained using  the solution 'a9'. We present the nominal past K2 orbit evolution in Fig. ~\ref{fig:2017k2a9_nomial}. One can notice a series of moderate stellar perturbations and a very strong perturbation caused by P0230 2.47~Myr ago. This solution gives the $1/a_{\rm ori}=35.48\pm2.33 \times 10^{-6}$au$^{-1}$, which corresponds to the previous orbital period of 4.73~Myr. As a result, the indirect perturbation from P0230 happened almost exactly when the comet was at its aphelion and moved very slowly. 

\begin{figure}
	\includegraphics[angle=270,width=1\columnwidth]{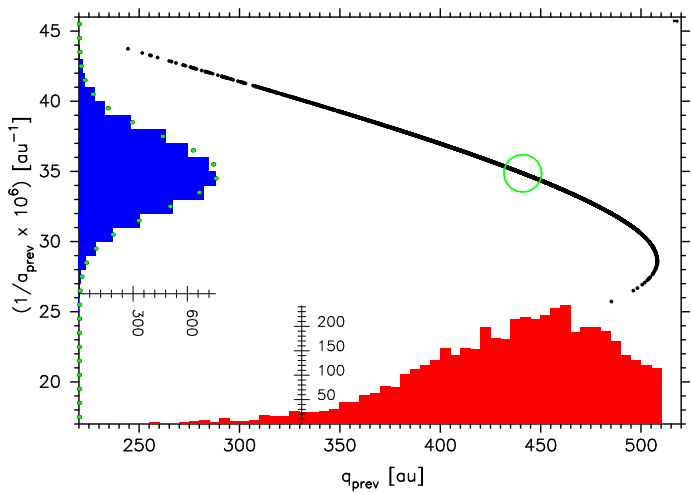} 
	\caption{ Influence of K2 orbit uncertainty (for solution 'a9') on the parameters of its previous orbit. The number of points, color and symbol meanings are the same as in Fig.~\ref{fig-2014unb8_hist}.}\label{fig-2017k2a9_hist2}
\end{figure}

In Fig.~\ref{fig:2017k2a9_nomial} local orbit changes due to the action of P0509 at $-0.55$~Myr and P0506 at $-1.08$~Myr are visible, as in the case of BB. In addition, small orbit changes resulting from a passage of P0508, 0.98~Myr ago, and P0417, 1.47~Myr ago, can also be observed. More information about these stellar perturbers can be found in Table~\ref{tab:star-list}. The accumulated stellar perturbations increased the nominal K2 previous perihelion distance from 12~au obtained when only Galactic potential is taken into account (thin lines in Fig.~\ref{fig:2017k2a9_nomial}) up to 441~au. We have checked numerically that in the absence of P0230 all other stellar perturbations cancel each other and the resulting previous perihelion distance would almost not be affected by stars. 

As was already shown, the nominal previous perihelion is only a highly approximate qualitative result. It must be checked what is the possible influence of cometary and stellar data uncertainties on the final result. The influence of  the uncertainty of the K2 orbit (for  the solution 'a9') on our results is presented in Fig.~\ref{fig-2017k2a9_hist2}. We used the methods described in Sect.~\ref{sect:uncertainties}. The  numerical results of this calculation are included in Table~\ref{tab:K2_previous_part1}. In Fig.~\ref{fig-2017k2a9_hist2} we also present marginal distributions of $1/a_{\rm prev}$ (blue) and $q_{\rm prev}$ (red). As the best estimate, we obtained from these calculations:  $1/a_{\rm prev} = (34.88 \pm 2.61) \times 10^{-6}$~au$^{-1}$ and  $q_{\rm prev}$ = (373.5--441.6--490.0)~au. The  latter  is in the form of three deciles, since its distribution is not Gaussian. It is easy to note that the range of the previous perihelion distance is well defined here, and its nominal and median values are close to each other. The similar numerical experiments we performed for the remaining orbital solutions and the results can be reviewed in Table~\ref{tab:K2_previous_part1}.

\begin{figure}
	\includegraphics[angle=270,width=1\columnwidth]{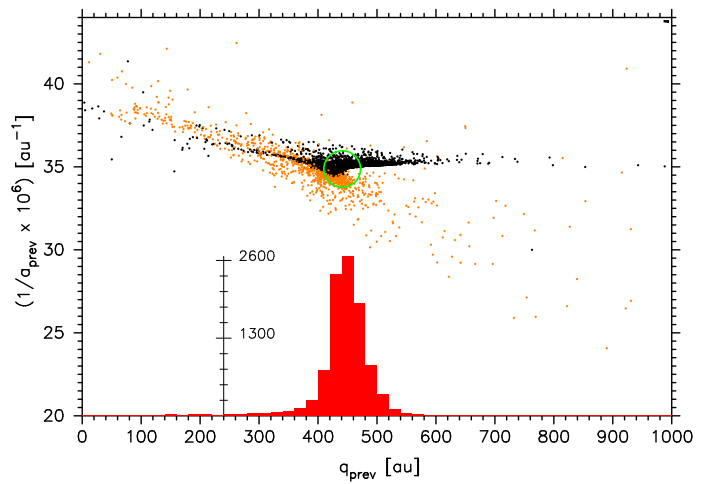} 
	\caption{ Influence of P0509 data uncertainty on the parameters of the previous K2 nominal orbit (solution 'a9'). Black dots mark 8827 cases where the distance between K2 and P0509 is greater than 20\,000~au while 1149 orange dots shows the opposite cases. Two black and 23 orange points are omitted as the extreme outliers: the maximum $q_{\rm prev}$ in this simulation equals 9510~au. The red marginal $q_{\rm prev}$ distribution was made for all 9976 plotted points.}\label{fig-2017k2a9_509summ}
\end{figure}

From Fig.~\ref{fig:2017k2a9_nomial} one can read that during the K2 backward numerical integration, from the listed stars, we first meet P0509 (star A in Fig.~\ref{fig-seven-stars}). This star data uncertainties are rather large. It has no parallax and proper motion data in {\it Gaia} EDR3, probably making some trouble in the observational data reduction, so we use all the astrometry for this star from DR2. We decided to check how important  this star might be  in our calculations. We draw 10\,000 clones of P0509 to observe the  spread of the result when calculating past K2 orbit parameters at the previous perihelion. The smallest star--comet distance in this experiment was 765~au. Results of this simulation are presented in Fig.~\ref{fig-2017k2a9_509summ} and Table~\ref{tab:K2_previous_part2} as K2-prev-C. Since this stellar perturbation is direct,  that is, P0509 acts on a comet strongly because it passes relatively close to it, we also analyzed a star--comet distance distribution. We use two different colors for dots representing the individual clone results: orange dots for clones with a minimal distance from K2 smaller than 20\,000~au and black for more distant ones. It can be noticed in Fig.~\ref{fig-2017k2a9_509summ} that the dependence between the strength of the perturbation and the star--comet distance is not obvious because it also strongly depends on the star--comet--Sun geometry. Moreover, some P0509 clones passed much closer to the Sun than the nominal one, therefore, an indirect perturbation is also possible. We omitted 15 cases as the extreme outliers. The smallest $q_{\rm prev}$ equals 3.6~au while the omitted largest one equals 9510~au. Despite such a large spread, the $q_{\rm prev}$ distribution shown as a red histogram in Fig.~\ref{fig-2017k2a9_509summ} is quite compact. It can be described by three deciles as: (407.1--447.6--489.1)~au. Still, we are close to the nominal value of $q_{\rm prev}$. 

\begin{figure}
	\includegraphics[angle=270,width=1\columnwidth]{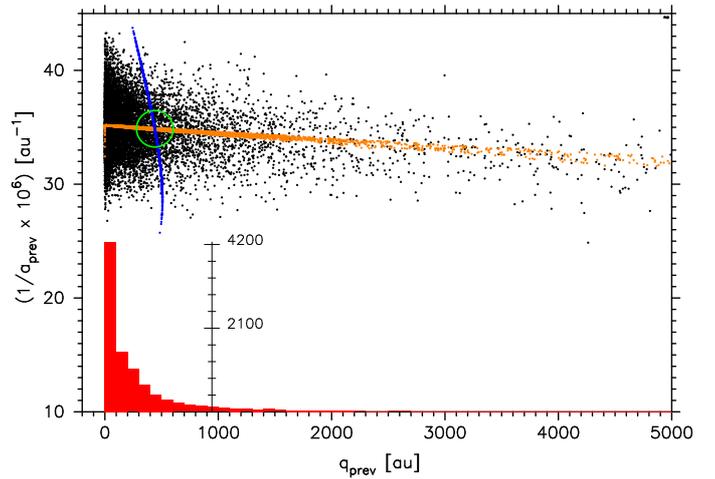} 
	\caption{ Composite picture of the influence of cometary and stellar uncertainties on the K2 orbit recorded at the previous perihelion passage. Blue dots mark 5001 K2 clones perturbed by all stars on their nominal tracks. Orange dots represent the nominal K2 orbit perturbed by  10\,000 P0230 clones. Black dots presents 10\,000 pairs of K2 and P0230 clones. Red marginal distribution histogram of $q_{\rm prev}$ is performed for the black cloud of points. The result for nominal stellar and cometary data is in the center of the green circle. Omitted are 226 outlier points from the far right tail of the black swarm and 219 from the orange one.}\label{fig-2017k2a9_230summ}
\end{figure}

The situation is quite different when we investigate the impact of  the uncertainties of P0230 on our results. We performed two additional simulations using the techniques presented in Sect.~\ref{sect:uncertainties} -- K2-prev-D: a comet in nominal orbit and 10\,000 clones of P0230, and K2-prev-E: 10\,000 pairs of a star clone and a comet clone. Numerical results of all simulations performed for K2, also for orbital solutions other than  'a9', are presented in Tabs.~\ref{tab:K2_previous_part1} and \ref{tab:K2_previous_part2}. The results based on the 'a9' solution are summarized in a composite plot shown in Fig.~\ref{fig-2017k2a9_230summ}. We plotted the results of K2-prev-E as black dots overprinted with orange dots from K2-prev-D. For the sake of comparison, we added the results of K2-prev-C (the same as presented in Fig.~\ref{fig-2017k2a9_hist2}) as blue dots on top of the previous two sets.

In Fig.~\ref{fig-2017k2a9_230summ} several interesting features can be seen. The K2 orbit uncertainties are responsible mainly for the spread in $1/a_{\rm prev}$ (blue dots). The swarm of P0230 clones, when acting on a nominal K2 orbit, causes mainly the spread in $q_{\rm prev}$ (orange dots) across a wide range of values from 0.0005~au to 19\,350.4~au. The third simulation, K2-prev-E (black dots), spread results in both elements. We observe here  a similar effect to that in  in Fig.~\ref{fig-2014unb8_230summ} -- the maximum of the $q_{\rm prev}$ distribution occurs for a substantially smaller value of the previous perihelion distance than that for the nominal orbit. The reason is also the same: due to the geometry of the P0230 swarm of random clones, the nominal star--Sun distance is much smaller than the most probable value. As a result, the weaker perturbation is more frequent than that for the nominal orbit of K2 and a star track. 

Despite the general indirect nature of the the influence of P0230 on K2, there are several star clones in K2-prev-E that pass close to K2. The minimum star--comet distance in this numerical experiment is 7707~au. Such close passages produce outlier results with negative $1/a_{\rm prev}$ values; the smallest value is $-621$ in the units used in the plot. When preparing Fig.~\ref{fig-2017k2a9_230summ} we omitted 226 points, 27 of them with negative $1/a_{\rm prev}$. The largest value of the right tail of the black swarm is $q_{\rm prev}=59\,312.6$~au.

Based  on the 'a9' solution, the K2-prev-E simulation presents the best we can say on the previous K2 orbit at the time of this writing. We obtained $q_{\rm prev}$ described by deciles: (5.163--144.0--1065)~au, which suggests that, contrary to our previous, opinion, K2 is probably a dynamically new comet. However, taking into account how sensitive this result is to the P0230 data change, we should wait for much more precise measurements for this star before any definitive statements can be formulated. 

\section{Summary and Conclusions}
\label{sect:summary}
We performed an extensive study of the past and future motion of comet C/2014~UN271  (Bernardinelli-Bernstein) and the past motion of C/2017~K2  (PANSTARRS). 
For each comet, we obtain a series of osculating orbits based on data arcs of different lengths.  In the case of BB,  we did not find detectable effects related to the NG~acceleration caused by the sublimation of ices from the comet surface. In K2 motion, measurable effects of NG~acceleration have  now been noticed for the data arc covering large heliocentric distances down to the 7.8~au (solution c5 in Table~\ref{tab:positional-data}). From all orbits listed in the above-quoted table, we  have selected the most appropriate ones for a dynamical study of each comet motion outside the planetary zone. In the case of BB, it is the solution 'b8', based on all observations available at the time of calculation (October 2021), for K2 -- the solution 'a9' based on the data arc shorter than the currently available. However, we also performed simulations for all remaining solutions given in Table~\ref{tab:positional-data} for comparison purposes.

The analysis of a series of K2 orbital solutions based on different data arcs clearly shows that in the case of LPCs discovered at large heliocentric distances, the starting orbits for the study of their origin should be determined from the pre-perihelion leg of the orbit and limited to large distances from the Sun. Moreover, in cases  such as K2, it is necessary to individually balance the benefits of limiting the action of NG~acceleration (the farther from the Sun the better) and the quality of the obtained orbit (the longer the data arc the better). This approach, extended to LPCs discovered at very large heliocentric distances, confirms our previous published conclusions \citep{kroli-dyb:2012, krolikowska:2020}.

Based  on the selected preferred orbit solutions, we obtained a nominal previous perihelion distance value. For K2, nominal $q_{\rm prev}=441$~au and for BB nominal $q_{\rm prev}=237$~au, taking into account the perturbing effect of the full Galactic tidal field and all currently known potential stellar perturbers. The influence of orbital uncertainties of the preferred comet orbital solutions is quite small in both cases and resulted in  $q_{\rm prev}=$(373.5--441.6--490.0)~au for K2 (deciles 10$^{\rm th}$, median, and 90$^{\rm th}$ are used here) and  $q_{\rm prev}=237.6\pm 32.9$~au for BB.  At this stage, we could conclude that both K2 and BB are dynamically new comets. However, as was shown above, the influence of the stellar data uncertainties on our results must be considered. For the K2 past motion, we separately analyze the effects caused by stars P0509 and P0230. In the case of BB past motion, we also experiment with P0230 and for  the BB future motion, we analyzed the effect of P0107 uncertainties.  Details of the above-mentioned results and several numerical experiments with the stellar perturbations can be found in Tables~\ref{tab:BB_previous_next}, \ref{tab:K2_previous_part1}, and \ref{tab:K2_previous_part2} and the corresponding figures.

The list of stars that noticeably change BB or K2 orbit evolution is presented in Table~\ref{tab:star-list} together with their parameters of the close Sun--star approaches and their mass estimates, based on the release 3.1 of the StePPeD database. The past motion of both studied comets is dominated by a perturbation caused by the star P0230 (\object{HD~7977}). Our most advanced simulations K2-prev-E and BB-prev-D resulted in significant changes in the previous perihelion distance values obtained:  $q_{\rm prev}=$~(5.163--144.0--1065.)~au for K2  and  $q_{\rm prev}=$~(12.50--49.62--252.4)~au for BB. Looking at the median values, we can still classify both comets as dynamically new. However, the observed high sensitivity of these results to the uncertainty of the P0230 data causes this conclusion to be preliminary and uncertain. We hope to receive  more precise data for P0230. 

It is worth mentioning that due to the specific nature of perturbations caused by P0107 and P0230, they can be the significant perturbers of many LPCs.

It is worth mentioning that due to the specific nature of perturbations caused by P0107 and P0230 they can be the significant perturbers of many LPCs. This effect comes from their very close passage near the Sun, which generates a strong velocity impulse indirectly affecting all Solar System bodies. As a result, many heliocentric small body orbits have been or will be changed. The strength of this perturbation depends mainly on the small body heliocentric velocity. Therefore, LPCs at their aphelia are the best candidates to track important changes in their orbits.

Our numerical Monte Carlo simulations of the impact of stellar data uncertainties on the long-term evolution of comet orbits clearly show that the accuracy of the stellar data is still insufficient in many cases.

Our numerical Monte Carlo simulations of the impact of stellar data uncertainties on a long-term evolution of comet orbits clearly show that  the accuracy of the stellar data is still insufficient in many cases.  On the  contrary, the accuracy of the orbits of the contemporary observed comets seems to be satisfactory. From a list of seven stars that significantly perturb motion of the two comets studied here, only for two we have satisfying 6D data. We hope that future observational attempts, for example, next {\it Gaia} data release, will improve the situation.

\begin{acknowledgements}
This research has made use of positional data of analyzed comets provided by the International Astronomical Union's Minor Planet Center. This research has also made use of the SIMBAD database, operated at CDS, Strasbourg, France, and the VizieR  catalog access tool, CDS, Strasbourg, France (DOI: 10.26093/cds/vizier). This work has made use of data from the European Space Agency (ESA) mission {\it Gaia} (\url{https://www.cosmos.esa.int/gaia}), processed by the {\it Gaia} Data Processing and Analysis Consortium (DPAC, \url{https://www.cosmos.esa.int/web/gaia/dpac/consortium}). Funding for the DPAC has been provided by national institutions, in particular the institutions participating in the {\it Gaia} Multilateral Agreement. The calculations which led to this work were in some part performed with the support from the project “GAVIP-GC: processing resources for Gaia data analysis” funded by European Space Agency (4000120180/17/NL/CBi).
\end{acknowledgements}

\bibliographystyle{aa}   
\bibliography{PAD31} 

\onecolumn

\begin{appendix}

\section{Orbital data for C/2014 UN$_{271}$ (Bernardinelli-Bernstein) \label{Ap:orbit_UN271}}

\begin{table*}[h]
\caption{Original barycentric orbits of C/2014 UN$_{271}$ at 250~au from the Sun , which was used as the starting orbit for the dynamical evolution discussed in this paper; 
}\label{tab:UN_original_orbits}
\setlength{\tabcolsep}{8pt} 
\centering
\begin{tabular}{lcc}
\hline
\hline 
 &  &  \\
solution                                   &  i1                        &  b8                                 \\
 &  &  \\
\hline 
 &  &  \\
time of the perihelion passage [TT]            & 2031\,01\,21.46447114 $\pm$ 0.018088 & 2031\,01\,21.48844110 $\pm$ 0.012463 \\
perihelion distance [au]                   & 10.94876759 $\pm$ .00010670          &  10.94877654 $\pm$ .00007315         \\
eccentricity                               &  0.99945070 $\pm$ .00000821          &   0.99943996 $\pm$ .00000616         \\
 inverse of the semimajor axis [$10^{-6}$~au$^{-1}$]            & 50.17 $\pm$ 0.75                     & 51.15 $\pm$ 0.56                     \\ 
inclination [deg]                          &  95.466167 $\pm$  0.000035         &  95.466152 $\pm$  0.000023          \\
argument of perihelion [deg]               & 326.280703 $\pm$  0.000515         & 326.280949 $\pm$  0.000363          \\
longitude of the ascending node [deg]      & 190.002743 $\pm$  0.000038         & 190.002732 $\pm$  0.000025          \\
epoch of osculation [TT]                   &  1715\,03\,12                      & 1715\,03\,12                       \\
\hline 
\end{tabular}
\end{table*}

\begin{table*}[h]
\caption{Future barycentric orbits of C/2014 UN$_{271}$ at 250~au from the sun, which was used as the starting orbit for the future dynamical evolution discussed in this paper. }\label{tab:UN_future_orbits}
\setlength{\tabcolsep}{8pt} 
\centering
\begin{tabular}{lcc}
\hline
\hline 
 &  &  \\
solution                                   &  i1                        &   b8                                 \\
 &  &  \\
\hline 
 &  &  \\
time of the perihelion passage [TT]            & 2031\,01\,21.43420845 $\pm$ 0.018081 & 2031\,01\,21.45818806 $\pm$ 0.012461 \\
perihelion distance [au]                   & 10.94853286 $\pm$ .00010668          &  10.94854181 $\pm$ .00007313         \\
eccentricity                               &  0.99960500 $\pm$ .00000821          &   0.99959425 $\pm$ .00000615         \\
inverse of the semimajor axis [$10^{-6}$~au$^{-1}$] & 36.08 $\pm$ 0.75                     & 37.06 $\pm$ 0.56                     \\ 
inclination [deg]                          &  95.460461 $\pm$  0.000035         &  95.460446 $\pm$  0.000023             \\
argument of perihelion [deg]               & 326.246713 $\pm$  0.000514         & 326.246959 $\pm$  0.000363             \\
longitude of the ascending node [deg]      & 190.009262 $\pm$  0.000038         & 190.009251 $\pm$  0.000025             \\
epoch of osculation [TT]                   &  2346\,08\,31                      & 2346\,10\,10                           \\
\hline 
\end{tabular}
\end{table*}

\section{Orbital data for C/2017 K2  (PANSTARRS) \label{Ap:orbit_K2}}

\begin{table*}[h]
\caption{Original barycentric orbits of C/2017 K2 at 250~au from the Sun, which were used as the starting orbit for the dynamical evolution discussed in this paper. } \label{tab:K2_original_orbits}
\setlength{\tabcolsep}{4.0pt} 
\centering
\begin{tabular}{lccc}
\hline
\hline 
 &  &  & \\
solution                                   &  a6                     & c5                           &  b5           \\
 &  &  & \\
\hline 
 &  &  & \\
time of perihelion passage ~~~~~TT = 2022       & 12\,18.803948 $\pm$ 0.000814    & 12\,18.793820 $\pm$ 0.002840 & 12\,18.814958 $\pm$ 0.000210 \\
perihelion distance [au]                   & 1.79565692 $\pm$ 0.00000240      & 1.79558257 $\pm$ 0.00000609        & 1.79561076 $\pm$ 0.00000124        \\
eccentricity                               & 0.99994071 $\pm$ .00000082       & 0.99993541 $\pm$ 0.00000163        & 0.99992319 $\pm$ 0.00000041        \\
inverse of the semimajor axis [$10^{-6}$~au$^{-1}$]              & 33.02 $\pm$ 0.45                  & 35.97 $\pm$ 0.97                   & 42.78 $\pm$ 0.23                   \\ 
inclination [deg]                          &  87.573120 $\pm$  0.000005       &  87.573189  $\pm$ 0.000077         &  87.573070 $\pm$  0.000003         \\
argument of perihelion [deg]               & 236.210369 $\pm$  0.000072       & 236.211575  $\pm$ 0.000161         & 236.212047 $\pm$  0.000035         \\
longitude of the ascending node [deg]      &  88.086676 $\pm$  0.000017       &  88.086242  $\pm$ 0.000051         &  88.086882 $\pm$  0.000008         \\
epoch of osculation [TT]                   & 1723\,01\,29                     & 1722\,12\,20                       & 1722\,12\,20              \\
\hline 
\end{tabular}
\tablefoot{
Original orbits representing solutions a8 and a9 were published in KD18a and KD18b, respectively.
}

\end{table*}
\vfill
\end{appendix}

\end{document}